\documentclass[twocolumn,letterpaper,prb,superscriptaddress,showpacs,amsmath]{revtex4}

\usepackage{graphicx}

\begin{document}

\title{Evidence for a superconducting origin of prominent features of the in-plane infrared response of underdoped cuprates
and implications of their persistence above {\boldmath $T_{\mathrm{c}}$}}

\author{B\v{r}etislav \v{S}op{\'\i}k}
\affiliation{Central European Institute of Technology, 
Masaryk University, Kotl\'a\v{r}sk\'a 2, 61137 Brno, Czech Republic}
\author{Ji\v{r}\'{\i} Chaloupka}
\affiliation{Central European Institute of Technology, 
Masaryk University, Kotl\'a\v{r}sk\'a 2, 61137 Brno, Czech Republic}
\author{Dominik Munzar}
\affiliation{Central European Institute of Technology, 
Masaryk University, Kotl\'a\v{r}sk\'a 2, 61137 Brno, Czech Republic}
\affiliation{Department of Condensed Matter Physics, Faculty of Science,  
Masaryk University, Kotl\'a\v{r}sk\'a 2, 61137 Brno, Czech Republic}

\begin{abstract}
We report on results of our analysis of published experimental data 
of the in-plane infrared response  
of two representative underdoped  high-$T_{\mathrm{c}}$ cuprate superconductors,     
focusing on a characteristic gap feature in the spectra of the real part of the conductivity 
and the corresponding structures of the memory function,   
that develop below a temperature $T^{\mathrm{ons}}$ considerably higher than $T_{\mathrm{c}}$. 
Several arguments based on comparisons of the data 
with results of our calculations are provided 
indicating that the features are due to superconductivity 
and that $T^{\mathrm{ons}}$ marks the onset of a precursor superconducting phase. 
\end{abstract}

\date{\today}

\maketitle 
The possible persistence of some form of superconductivity 
many tens of ${\mathrm K}$ above the~bulk superconducting transition temperature $T_{\mathrm{c}}$ 
in underdoped cuprate superconductors 
belongs to the most vividly discussed topics 
in the field of high-$T_{\mathrm{c}}$ superconductivity, 
for representative examples of related experimental studies,  see Refs.~\onlinecite{Wang:2006:PRB,
Li:2010:PRB,
Li:2013:PRB,
Gomes:2007:Science,
Dubroka:2011:PRL,
Kaiser:2012:condmat, 
Kondo:2011:NaturePhys,
Reber:2012:NaturePhys,
Reber:2013:PRB,
Tallon:2013:PRB,
Mahyari:2013:PRB}. 
Surprisingly high (up to $\approx 100\mathrm{\,K}$ above $T_{\mathrm{c}}$) values 
of the temperature $T^{\mathrm{ons}}$ 
of the onset of an increase of coherence, 
presumably due to an onset of a precursor superconducting phase, 
have been deduced from the data of the $c$-axis infrared response 
of underdoped YBa$_{2}$Cu$_{3}$O$_{7-\delta}$ (Y-123)\cite{Dubroka:2011:PRL}. 
The interpretation of the $T^{\mathrm{ons}}$ scale in terms of a precursor superconductivity, however, 
has not yet been widely accepted. 
The main reasons are: 
(i) The $c$-axis response of Y-123 is a fairly complex quantity 
due to the specific bilayer structure of this compound.  
(ii) Underdoped cuprates are known to exhibit ordered states distinct from superconductivity, 
in particular, charge modulations have been reported\cite{Ghiringhelli:2012:Science,Chang:2012:NaturePhys}  
that set on at temperatures comparable to $T^{\mathrm{ons}}$.
It is thus possible to speculate that the $T^{\mathrm{ons}}$ scale is determined by an order competing with superconductivity 
rather than by superconducting correlations themselves.  
In this context it is of high importance to identify 
manifestations of the increase of coherence below $T^{\mathrm{ons}}$ in the in-plane response,    
a quantity, that is less sensitive to structural details than the $c$-axis response,  
and to ascertain their relation to superconductivity.  

It has been already shown by Dubroka {\it et al.}\cite{Dubroka:2011:PRL} 
that the in-plane infrared conductivity 
of underdoped Y-123 
changes at $T^{\mathrm{ons}}$ 
in a way similar to that of an optimally doped  superconductor at $T_{\mathrm{c}}$.  
The focus, however, has been on qualitative aspects of the relevant spectral weight shifts,  
and the related spectral structures have not been addressed. 
Here we concentrate on the temperature dependence of three prominent spectral features 
that develop below $T^{\mathrm{ons}}$.   
Our analysis involves comparisons of the data of two representative underdoped cuprates 
with those of optimally doped ones and   
with~results of our calculations employing approaches
ranging from the Allen's theory to~the fully selfconsistent generalized Eliashberg theory. 
It provides evidence that the features are due to superconductivity,  
and the scale $T^{\mathrm{ons}}$ due to a form of superconductivity 
rather than to an ordered state competing with superconductivity.

The paper is organized as follows. 
First, we summarize the relevant aspects of the experimental infrared data. 
As examples we use the published data 
of underdoped (UD) HgBa$_{2}$CuO$_{4+\delta}$ (Hg-1201)\cite{Mirzaei:2013:PNAS}
and of underdoped Y-123\cite{Hwang:2006:PRB,Dubroka:2011:PRL}. 
Next we compare them with the published data of optimally doped (OPD) cuprates 
and, in particular, with the spectra calculated using the Eliashberg theory. 
We put emphasis on the similarity between the low temperature ($T$) data and the low-$T$ calculated spectra
and on the similarity between the onset of the features below $T^{\mathrm{ons}}$ in UD cuprates 
and that below $T_{\mathrm{c}}$ in the calculated spectra. 
In the next paragraph, the data are discussed in terms of the frequently used extended Allen's theory (EAT)
\cite{Sharapov:2005:PRB,Hwang:2006:PRB,Hwang:2008:PRL,Hwang:2011:PRB}.    
It will be highlighted     
that the assumption of a superconductivity unrelated gap in the density of states (DOS) 
in the  temperature range $T_{\mathrm{c}}<T<T^{\mathrm{ons}}$
leads to clear inconsistencies with the data.
Finally, we present and discuss results of our calculations of the optical spectra, 
where the gap in the DOS is of superconducting origin, 
using as inputs recently published properties of the quasiparticle spectral function 
obtained from the photoemission data by means of the tomographic density of states method\cite{Reber:2012:NaturePhys}.  

{\it Relevant aspects of the experimental data}. 
The gap feature in the spectra of the real part $\sigma_{1}$ of the infrared conductivity $\sigma$ 
and the related structures of the memory function $M(\omega)=M_{1}(\omega)+M_{2}(\omega)$  
defined by $\sigma(\omega)=i\varepsilon_{0}\omega_{\mathrm{pl}}^{2}/[M(\omega)+\omega]$\cite{Gotze:1972:PRB,Optselfenergy}, 
will be illustrated 
with the recently published data of UD Hg-1201 with $10\%$ doping 
and $T_{\mathrm{c}}=67\mathrm{\,K}$\cite{Mirzaei:2013:PNAS}. 
The data are similar to the earlier published ones of comparably UD Y-123\cite{Hwang:2006:PRB,Dubroka:2011:PRL},  
but a much more detailed temperature dependence is reported and
some spectral features are sharper than in Y-123. 
Figure 1 shows, in part (a), a selection of the spectra of $\sigma_{1}$ of Hg-1201 from Fig.~1 of Ref.~\onlinecite{Mirzaei:2013:PNAS}, 
and, in parts (b) and (c), the corresponding spectra of $M_{1}$ and $M_{2}$ 
from Fig.~4 of Ref.~\onlinecite{Mirzaei:2013:PNAS}.
\begin{figure*}[tbp]
\includegraphics{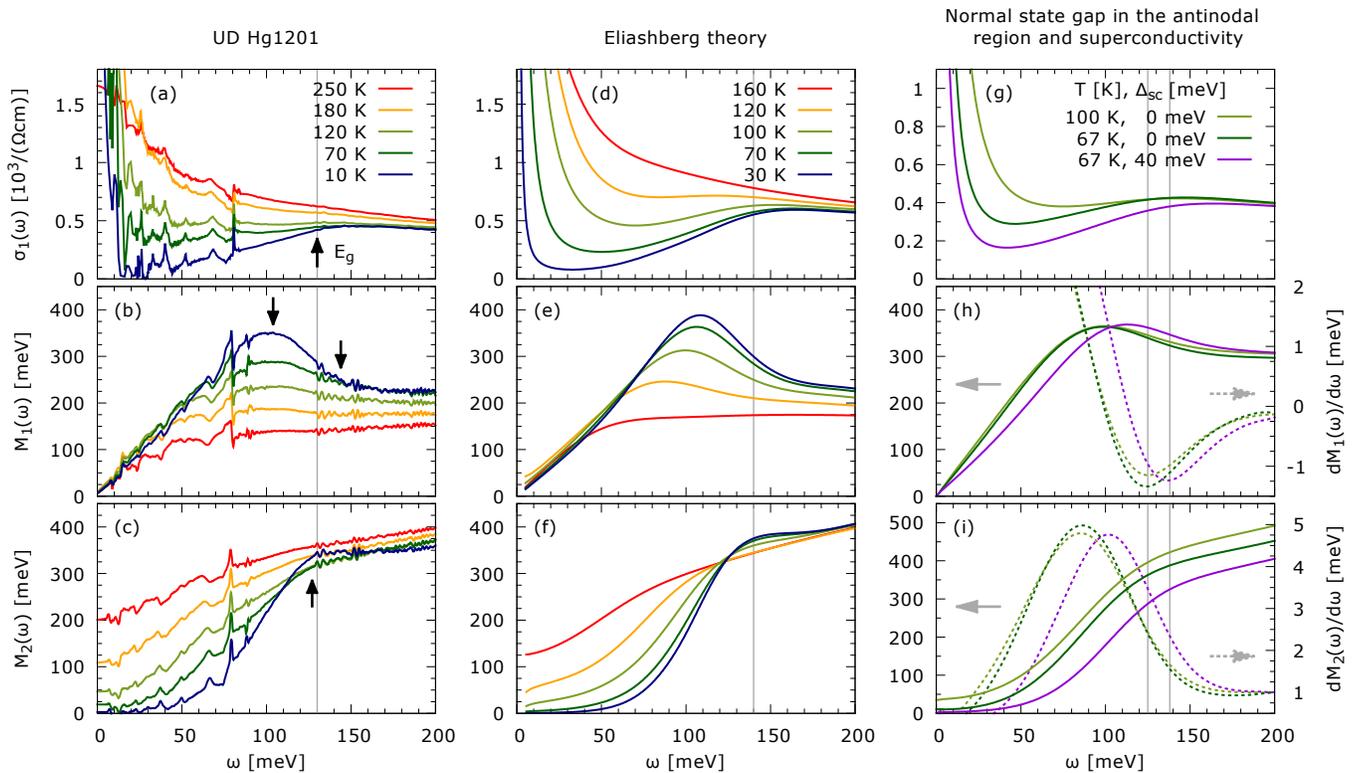}
\caption{(a) Real part of the in-plane infrared conductivity  
of Hg-1201 with $10\%$ doping and $T_{c}=67\mathrm{\,K}$ for five selected temperatures 
(data obtained by Mirzaei and coworkers, extracted from Fig.~1 of Ref.~\onlinecite{Mirzaei:2013:PNAS}). 
(b) and (c): The corresponding spectra of $M_{1}$ and $M_{2}$, respectively
(data from Fig.~4 of Ref.~\onlinecite{Mirzaei:2013:PNAS}).
The arrows indicate the features discussed in the text:  
the one in (a) the change of slope of $\sigma_{1}$,   
the left (right) one in (b) the maximum (the onset feature at the high energy side) 
of the characteristic peak of $M_{1}$,  
and that in (c) indicates the kink of $M_{2}$.  
(d), (e) and (f): The spectra of $\sigma_{1}$, $M_{1}$ and $M_{2}$ 
calculated using the model of charged quasiparticles coupled to spin fluctuations 
and the fully selfconsistent generalized Eliashberg equations, 
for details see Sec.~I of Ref.~\onlinecite{Supportinginformation}.   
(g), (h), and (i): The spectra of $\sigma_{1}$, $M_{1}$ and $M_{2}$ calculated using the hybrid approach,  
for details see Sec.~IV of Ref.~\onlinecite{Supportinginformation}.   
The dotted lines in (h) and (i) represent the derivatives of $M_{1}$ and $M_{2}$. 
They are shown to highlight the shift of the structures due to $\Delta_{\mathrm{SC}}$. 
}
\label{fig:Expdata}
\end{figure*}
The low temperature spectra of $\sigma_{1}$ exhibit a gap ranging from low energies to 
$E_{\mathrm{g}}\approx 130\mathrm{\,meV}$, 
where a maximum occurs and the slope of the spectra changes, see the arrow in Fig.~1~(a).  
The change of the slope will be called the gap edge in the following. 
It is accompanied by a characteristic peak in the spectra of $M_{1}$ 
with an onset (change of the slope) feature at the high energy side,  
and a kink in those of $M_{2}$, see the arrows in Figs.~1~(b) and (c). 
All the three features are well known from earlier infrared studies\cite{Timusk:1999:RPP,
Basov:2005:RMP,
Hwang:2006:PRB,
Hwang:2008:PRL, 
Hwang:2011:PRB}. 
What has, however,---to the best of our knowledge---not yet been recognized, is the fact
that they appear close to the temperature $T^{\mathrm{ons}}$ of Ref.~\onlinecite{Dubroka:2011:PRL}. 
This will be discussed below. 
We focus on $\sigma_{1}$ first. 
The $70\mathrm{\,K}$ and $120\mathrm{\,K}$ spectra in Fig.~1 (a) display a clear gap edge at an energy close to $E_{\mathrm{g}}$.
The $250\mathrm{\,K}$ spectrum,  on the other hand, does not display any such feature in the relevant spectral range. 
It can be seen in Fig.~1 of Ref.~\onlinecite{Mirzaei:2013:PNAS}, that the latter vanishes between $180\mathrm{\,K}$ and $200\mathrm{\,K}$, 
close to the maximum $T^{\mathrm{ons}}$ of Ref.~\onlinecite{Dubroka:2011:PRL}. 
The temperature dependence of the characteristic peak in the spectra of $M_{1}$ is similar:
at $120\mathrm{\,K}$ it is still very clear and almost at the same location as at low temperatures, 
at $250\mathrm{\,K}$ it is absent.  
Note that the onset feature at the high energy side of a similar peak occuring in the case of UD Y-123 
has already been reported to vanish around $170\mathrm{\,K}$, see the discussion of Fig.~7 of Ref.~\onlinecite{Hwang:2006:PRB}. 
Next we address the temperature dependence of the kink in the spectra of $M_{2}$.
At low temperatures the feature is very sharp 
(note that the $10\mathrm{\,K}$ spectrum overshoots the $70\mathrm{\,K}$ one).  
It gets smoother slightly above $T_{\mathrm{c}}$,  persists, at approximately the same energy, up to much higher temperatures,  
and cannot be resolved for temperatures higher than ca $200\mathrm{\,K}$. 

{\it Comparison with the data of OPD cuprates.} 
In OPD materials, a gap feature in $\sigma_{1}(\omega)$, a peak in $M_{1}(\omega)$ 
and a kink in $M_{2}(\omega)$, 
similar to those discussed above, 
set on at $T_{\mathrm{c}}$ (or very slightly above $T_{\mathrm{c}}$), 
for representative examples, see Refs.~\onlinecite{Puchkov:1996:JPhysCM,Marel:2003:Nature,Boris:2004:Science,Hwang:2004:Nature}. 
These features of OPD materials are clearly caused by superconductivity, 
since they develop in parallel with the formation of the loss-free contribution to $\sigma(\omega)$ of the superconducting condensate. 
It appears, that the three features of OPD materials, that set on at $T_{\mathrm{c}}$ and are due to superconductivity, 
continuously transform into those of the UD ones, setting on at $T^{\mathrm{ons}}$. 
This is a strong phenomenological argument 
in favour of the superconductivity-based interpretation of the three features and 
the precursor-superconductivity based interpretation of the $T^{\mathrm{ons}}$ scale. 
Recall that the low-$T$ superconducting state spectra of $\sigma$ of OPD materials 
are well understood in terms of the model, 
where charged planar quasiparticles are coupled to spin fluctuations,
whose spectrum consists of the resonance mode and 
a continuum\cite{Munzar:1999:PhysicaC,Carbotte:1999:Nature,Abanov:2001:PRB,Casek:2005:PRB,Chaloupka:2007:PRB}. 
In the following paragraph we demonstrate that the same model provides low-$T$ spectra in quantitative agreement 
with the low-$T$ data of the underdoped Hg-1201. 

{\it Comparison with the calculated (Eliashberg) spectra.}
Figures 1 (d), (e) and (f) show the spectra of $\sigma_{1}$, $M_{1}$ and $M_{2}$, respectively, 
calculated using the model and a real axis version of the fully selfconsistent approach of Ref.~\onlinecite{Chaloupka:2007:PRB}. 
Computational details and the values of the input parameters, 
providing the values of $T_{\mathrm{c}}$ and $\Delta_{\mathrm{max}}$ of $133\mathrm{\,K}$ and $45\mathrm{\,meV}$, respectively,  
are given in Sec.~I of the supporting information, Ref.~\onlinecite{Supportinginformation}.
The high value of $T_{\mathrm{c}}$ is not surprising considering the observations of Ref.~\onlinecite{Dahm:2009:NaturePhys}.
Note the striking agreement between the low-$T$ spectra and the low-$T$ data shown in Figs.~1 (a), (b) and (c).
The structures of $\sigma_{1}$ and $M_{2}$ can be interpreted along the lines of the earlier theoretical 
studies\cite{Casek:2005:PRB,Chaloupka:2007:PRB}, 
for a summary, see Fig.~2. 
\begin{figure}[tbp]
\includegraphics{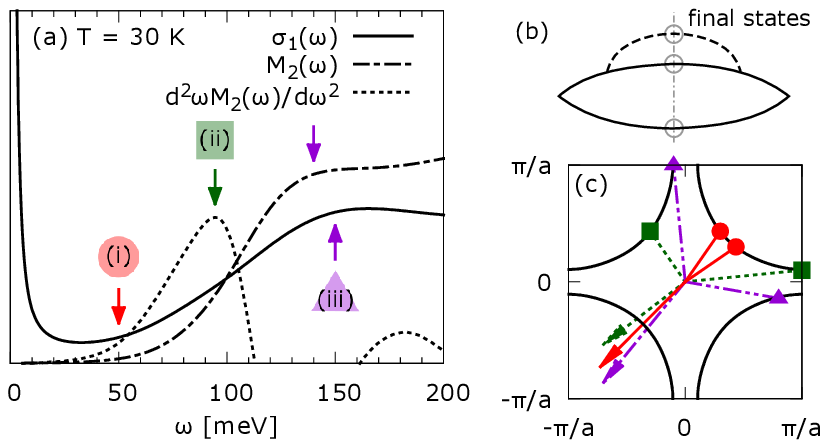}
\caption{(a) The $30\mathrm{\,K}$ spectra of $\sigma_{1}$ and $M_{2}$ from Figs.~1 (d) and (f)
and the corresponding spectra of the second derivative of $\omega M_{2}$, all in arbitrary units.
The arrows indicate three important features, occuring also in the experimental spectra 
of OPD cuprates, whose interpretation will be presented.  
(i) A weak onset of $\sigma_{1}$ and $M_{2}$ at the energy $\hbar\omega_{0}$ of the resonance
mode\cite{Casek:2005:PRB,Chaloupka:2007:PRB,Marel:2003:Nature,Boris:2004:Science}.
(ii) A maximum of the second derivative of $\omega M_{2}$\cite{Marsiglio:1998:PhysLettA} 
at approximately $\hbar\omega_{0}+\Delta_{\mathrm{max}}$\cite{Carbotte:1999:Nature,Chaloupka:2007:PRB},
where $\Delta_{\mathrm{max}}$ is the maximum of the superconducting gap of $d_{x^{2}-y^{2}}$ symmetry.
(iii) A well defined maximum of $\sigma_{1}$
at $E_{\mathrm{g}}$ with $E_{\mathrm{g}}$ approximately equal
or slightly above $\hbar\omega_{0}+2\Delta_{\mathrm{max}}$\cite{Chaloupka:2007:PRB,Marel:2003:Nature,Boris:2004:Science}, 
i.e., the gap edge discussed in the preceding paragraphs, 
and the corresponding kink of $M_{2}$. 
The features are due, as detailed in Refs.~\onlinecite{Casek:2005:PRB,Chaloupka:2007:PRB},
to the appearance above the characteristic energies of final states consisting of
(i) two near nodal (Bogolyubov) quasiparticles and the resonance,
(ii) a near nodal quasiparticle, an antinodal quasiparticle and the resonance, and
(iii) two antinodal quasiparticles and the resonance.
Above $\hbar\omega_{0}+2\Delta_{\mathrm{max}}$, the density of available final states saturates.
(b) Diagrammatic representation of final states discussed above.
The solid lines denote Bogolyubov quasiparticles, the dashed line the boson participating in the final state, 
within the present scheme the resonance. 
(c) A schematic representation of the three types of final states associated with the features (i), (ii), and (iii).
Shown is the first Brillouin zone and the Fermi surface. The symbols represent the Bogolyubov quasiparticles, 
the arrows the ${\mathbf k}$-vectors of the bosons.  
}
\label{fig:Schemeoffinalstates}
\end{figure}
The relation $E_{\mathrm{g}} \approx 2\Delta_{\mathrm{max}}+\hbar\omega_{0}$ (see the caption of Fig.~2) 
allows for a quantitative consistency check.  
The value of $E_{\mathrm{g}}$ of UD Hg-1201 of $130\mathrm{\,meV}$ is indeed approximately consistent 
with that of $\hbar\omega_{0}$ of $38\mathrm{\,meV}$ of UD Hg-1201 reported by Yuan Li {\it et al.}\cite{Li:2012:NaturePhys} 
and the value of $2\Delta_{\mathrm{max}}$ of $90\mathrm{\,meV}$ 
estimated based on the energy of the pairing peak in the Raman spectra of UD Hg-1201 of Ref.~\onlinecite{Li:2012:PRL}. 
To conclude, the low temperature infrared spectra of UD Hg-1201 are fully consistent 
with our model of a d-wave superconductor. 

Next we address the temperature dependence (TD) of the model spectra.  
It can be seen that the three important features---the gap feature in $\sigma_{1}$, 
the characteristic peak of $M_{1}$,  
and the sharp kink of $M_{2}$---develop 
below $T_{\mathrm{c}}$, in the same fashion as in the data of OPD 
materials\cite{Puchkov:1996:JPhysCM,Marel:2003:Nature,Boris:2004:Science,Hwang:2004:Nature}. 
In particular, the gap edge forms already close to $T_{\mathrm{c}}$.   
What we would like to highlight here is that not only the TD of the data of OPD cuprates 
but also that of the UD
is  similar to the model spectra, compare Figs.~1 (a), (b), (c) with Figs.~1 (d), (e), and (f). 
Note that the development of the gap edge in the UD Hg-1201  below $T^{\mathrm{ons}}\approx 200\mathrm{\,K}$ 
is analogous to that of the model spectra below $T_{\mathrm{c}}$: 
somewhat below $T^{\mathrm{ons}}$/$T_{\mathrm{c}}$  
a characteristic change of slope appears in $\sigma_{1}(\omega)$, 
with decreasing temperature its energy approaches $E_{\mathrm{g}}$ 
and a real gap and the maximum form. 
The similarity between the TD of the data below $T^{\mathrm{ons}}$ and that of the model spectra below $T_{\mathrm{c}}$ 
provides a further support for the interpretation of the $T^{\mathrm{ons}}$ scale in terms of a precursor superconductivity.

{\it Analysis of the above $T_{\mathrm{c}}$ data based on the EAT, 
problems of interpretations not involving superconductivity.}
The normal state (i.e., above $T_{\mathrm{c}}$) spectra of $M_{1}$ and $M_{2}$ 
of UD Y-123 and Bi$_{2}$Sr$_{2}$CaCu$_{2}$O$_{8+\delta}$ (Bi-2212) 
have been previously successfully fitted and interpreted
in terms of the phenomenological EAT\cite{Hwang:2008:PRL,Hwang:2011:PRB}.    
Note that the application of the EAT is based on two implicit assumptions:  
(i) The in-plane response is dominated by the contribution of near nodal quasiparticles;   
(ii) The dominant part of the nodal quasiparticle renormalization comes from a nodal-antinodal boson assisted scattering, 
the antinodal region being gapped.  
The essential inputs are 
the function $\alpha^{2}F$ describing bosonic excitations and  
the DOS displaying a gap at the Fermi energy, whose physical origin is not specified.    
Here we provide two indications that the gap captured by the EAT is likely related to superconductivity. 
(i) We have fitted the Hwang's data of Ref.~\onlinecite{Hwang:2006:PRB} using the formulas of the EAT 
and achieved a degree of agreement comparable to that of Ref.~\onlinecite{Hwang:2011:PRB}.  
Details can be found in Sec.~II of Ref.~\onlinecite{Supportinginformation}.  
The important points are: (a) While the presence of the DOS gap is needed 
to achieve a high quality fit for temperatures well below $T^{\mathrm{ons}}$, 
it is not essential for higher temperatures. 
For the latter (and not for the former, see Sec.~III of Ref.~\onlinecite{Supportinginformation}), 
the simple Allen's theory\cite{Allen:1971:PRB,Allen:2004:condmat}, 
as used in Refs.~\onlinecite{VanHeumen:2009:NJP,VanHeumen:2009:PRB},  appears to be sufficient.   
The temperature scale of the gap of the EAT based fits is thus $T^{\mathrm{ons}}$ rather than $T^{*}$;
(b) The opening of the gap in the DOS 
causes a spectral weight shift from the gap region to low frequencies, 
as expected for a precursor superconducting state. 
(ii) Assuming that the physics at $T_{\mathrm{c}}<T<T^{\mathrm{ons}}$ is unrelated to superconductivity, 
we arrive at a contradiction with the experimental data, as outlined below. 
Starting from the above assumption,   
the opening of the superconducting gap in the near nodal region below $T_{\mathrm{c}}$
can be expected to shift the structures established above $T_{\mathrm{c}}$ to higher energies. 
This shift indeed occurs in the calculated spectra but not in the experimental data. 
Figure 1 (g) shows the normal ($100\mathrm{\,K}$ and $67\mathrm{\,K}$) and the superconducting state ($67\mathrm{\,K}$) spectra of $\sigma_{1}$ 
calculated using the {\it hybrid approach}, described, e.g., in Ref.~\onlinecite{Casek:2005:PRB}, \onlinecite{Eschrig:2006:Advances},
employing the Nambu Green's functions and 
involving the (bare) dispersion relation of the charged quasiparticles, the superconducting gap $\Delta_{\mathbf k}$
($=0$ in the normal case) and the quasiparticle selfenergy.
The latter is taken from our EAT based fits of the Y-123 data of Ref.~\onlinecite{Hwang:2006:PRB}.  
For details of the calculations, see Sec.~IV of Ref.~\onlinecite{Supportinginformation}.        
The normal state, $67\mathrm{\,K}$ conductivity spectrum is very close to the experimental data and exhibits a gap edge around ca $125\mathrm{\,meV}$.  
In the superconducting state, the feature is clearly shifted to higher energies, and
the same applies to the spectra of $M_{1}$ and $M_{2}$ shown in Fig.~1 (h) and Fig.~1 (i). 
There is no such shift present in the data. 

{\it Comparison with the spectra calculated using results obtained by Reber and coworkers}. 
Reber {\it et al.}\cite{Reber:2012:NaturePhys,Reber:2013:PRB} have recently analyzed 
their photoemission data of OPD and UD Bi-2212
in terms of the tomographic density of states. 
This has led them to the observation that the near-nodal gap evolves smoothly through $T_{\mathrm{c}}$
and closes only at a temperature $T_{\mathrm{close}}$, that is for UD samples considerably higher than $T_{\mathrm{c}}$ (see Ref.~\onlinecite{Reber:2013:PRB}) 
and presumably close to $T^{\mathrm{ons}}$. 
This is a photoemission data based  evidence for the precursor superconductivity scenario.
Here we demonstrate that the energy scales of the infrared data and those of the photoemission ones are consistent with each other. 
We further argue that the data, taken together, imply the presence of superconducting correlations  
in the temperature range from $T_{\mathrm{c}}$ to $T^{\mathrm{ons}}$.    
Figure 3 shows the temperature dependence of $\sigma_{1}$ calculated using the hybrid approach  
but with the gap magnitude and the structure of the quasiparticle selfenergy from Ref.~\onlinecite{Reber:2012:NaturePhys}. 
Details can be found in Sec.~IV of Ref.~\onlinecite{Supportinginformation}. 
\begin{figure}[tbp]
\includegraphics{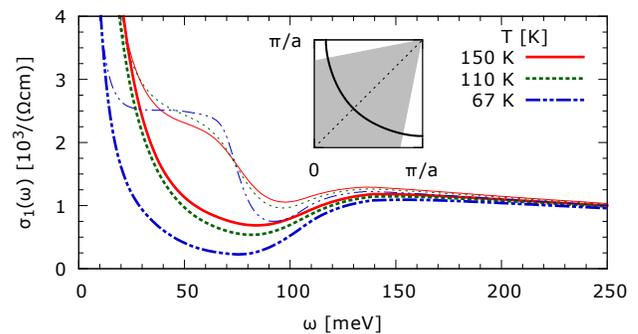}
\caption{Temperature dependence of $\sigma_{1}(\omega)$  
calculated using the hybrid approach and the gap magnitude and the structure of the quasiparticle selfenergy as in Ref.~\onlinecite{Reber:2012:NaturePhys}. 
Shown is the contribution of the region of the Brillouin-zone that is shaded in the inset. 
The thick (thin) lines represent results obtained 
with the off-diagonal component of the Green's function included, as in standard superconducting state calculations 
(not included, as in standard normal state calculations).
}
\label{fig:Sigma1fromresultsofReber}
\end{figure}
The thick (thin) lines represent the spectra calculated 
with the off-diagonal component of the Green's function included, as in standard superconducting state calculations 
(not included, as in standard normal state calculations).
First of all, it can be seen that the gap edge in $\sigma_{1}$ occurs 
at an energy very close to that of the experimental infrared data of underdoped cuprates. 
Second, the thin lines display a pronounced structure in the middle of the gap, 
whose energy is determined by $\Delta_{\mathrm{max}}$, here it is located at ca $60{\mathrm{\,meV}}$.     
It corresponds to transitions accross the superconducting gap, that are not allowed in the coherent superconducting state. 
The absence of the structure in the infrared data of UD Y-123 
(compare, e.g.,  the $67{\mathrm{\, K}}$ line in Fig.~3 of Ref.~\onlinecite{Hwang:2006:PRB} with the $100{\mathrm{\,K}}$ one) 
rules out the possibility that 
the above-$T_{\mathrm{c}}$ near nodal gap reported by Reber {\it et al.} 
and the related above $T_{\mathrm{c}}$ gap edge in $\sigma_{1}$
do not involve superconducting correlations. 

In conclusion, the published experimental data of $\sigma_{1}$ of underdoped high-$T_{\mathrm{c}}$ cuprate superconductors (HTCS)
display a clear gap feature below ca $130\mathrm{\,meV}$, 
setting on at $T^{\mathrm{ons}}>T_{\mathrm{c}}$. 
This is accompanied by the corresponding structures of the memory function. 
The features are similar to those of optimally doped HTCS setting on at $T_{\mathrm{c}}$, 
that are clearly due to superconductivity and well understood in terms of Eliashberg theory. 
This similarity and 
the one between the data and our calculated (Eliashberg) spectra 
strongly suggest that the gap feature of the underdoped HTCS
is also due to superconductivity and its persistence at $T_{\mathrm{c}}<T<T^{\mathrm{ons}}$ 
due to the presence of a precursor superconducting phase. 
In order to support this interpretation, we have demonstrated that  
(a) the temperature dependence of the feature 
cannot be simply accounted for in terms of a normal state gap independent of superconductivity and 
(b) the infrared data taken 
together with findings of recent photoemission studies employing the tomographic density of states method 
imply the presence of superconducting correlations in a broad range of temperatures above $T_{\mathrm{c}}$.   

This work was supported by the project CEITEC---Central European Institute of Technology (CZ.1.05/1.1.00/02.0068)
from European Regional Development Fund. 
B.~S.~was supported by the Program Nr.~CZ.1.07/2.3.00/30.0009 
and by the Metacentrum computing facilities(LM2010005). 
J.~Ch.~was supported by the AvH Foundation
and by EC 7$^\text{th}$ Framework Programme (286154/SYLICA).
Extensive discussions with A.~Dubroka and C.~Bernhard 
are gratefully acknowledged. 
We thank T.~Timusk for providing us the data of underdoped Y-123 
reported in Ref.~\onlinecite{Hwang:2006:PRB} and D.~Geffroy and A.~Dubroka 
for a critical reading of the manuscript.


\begin{thebibliography}{99}
\bibitem{Optselfenergy} The so called optical selfenergy $\sum^{\mathrm{opt}}$ 
is defined by $\sum^{\mathrm{opt}}(\omega)=-M(\omega)/2$.
\end{thebibliography}
\end{document}


\title{SUPPORTING MATERIAL \\ Evidence for superconducting origin of prominent features of the in-plane infrared response of underdoped cuprates and implications of their persistence above $T_{\rm c}$} 

\author{B\v{r}etislav \v{S}op{\'\i}k}
\author{Ji\v{r}\'{\i} Chaloupka}
\affiliation{Central European Institute of Technology, 
Masaryk University, Kotl\'a\v{r}sk\'a 2, 61137 Brno, Czech Republic}
\author{Dominik Munzar}
\affiliation{Central European Institute of Technology, 
Masaryk University, Kotl\'a\v{r}sk\'a 2, 61137 Brno, Czech Republic}
\affiliation{Department of Condensed Matter Physics, Faculty of Science, Masaryk University, Kotl{\' a}{\v r}sk{\' a} 2, 61137 Brno, Czech Republic}

\date{\today}
\pacs{74.25.Gz, 74.72.-h}

\maketitle 

\section{Generalized Eliashberg equations for the model of charged
quasiparticles coupled to spin fluctuations}
\label{sec:1}

Within the framework of the spin-fermion model used to obtain the data presented in Fig.~1~(d)--(f) of the main text, the superconductivity emerges in a similar way as in a coupled electron-phonon system corresponding to a conventional superconductor. The retarded pairing interaction is mediated by spin fluctuations replacing the phonons of the conventional case and coupling to the spin of the quasiparticles instead of their charge. The quantitative treatment of the spin-fermion model can be based on the generalized Eliashberg equations\cite{Eliashberg:1960:JETP,Eliashberg:1960:ZETF} with a modified coupling vertex and the phonon propagator replaced with the negatively taken spin susceptibility. The matrix selfenergy $\hat \Sigma$, in terms of Pauli matrices $\hat \Sigma = \Sigma^{(0)} \hat \tau_0 + \Sigma^{(3)} \hat \tau_3 + \Sigma^{(1)} \hat \tau_1$, is then determined by the selfconsistent equation 
\begin{equation}\label{eq:MatsselfE}
\hat \Sigma({\rm i}\omega_n,{\bf k}) = \frac{g^2}{\beta N} \sum_{m,{\bf q}} \chi({\rm i}\nu_m,{\bf q})\, \mathcal{\hat G}({\rm i}\omega_n-{\rm i}\nu_m,{\bf k}-{\bf q}),
\end{equation}
involving the renormalized quasiparticle propagator
\begin{equation}\label{eq:MatsGreen}
\mathcal{\hat G}^{-1}({\rm i}\omega_n,{\bf k})= [{\rm i}\omega_n-\Sigma^{(0)}({\rm i}\omega_n,{\bf k})] \hat \tau_0 -[\epsilon({\bf k})-\mu+\Sigma^{(3)}({\rm i}\omega_n,{\bf k})] \hat \tau_3 -\Sigma^{(1)}({\rm i}\omega_n,{\bf k}) \hat \tau_1
\end{equation}
and the spin susceptibility $\chi({\rm i}\nu_m,{\bf q})$, where ${\rm i}\omega_n$ are Matsubara energies for fermions and ${\rm i}\nu_m$ for bosons. The value of the coupling constant $g$ can be adjusted so that realistic values of the superconducting gap and of the transition temperature are obtained. 

A direct solution of Eq.~\eqref{eq:MatsselfE} in Matsubara energies has to be followed by a numerical analytical continuation of the selfenergy to the real axis which is an ill-posed problem. To avoid the numerical difficulties, one can employ the spectral representation of the quasiparticle propagator
\begin{equation}\label{eq:GreenSpectral}
\mathcal{\hat G}({\rm i}\omega_n, {\bf k})=\int_{-\infty}^\infty
\frac{\mathrm{d}\omega}{2\pi}\,\frac{\mathcal{\hat A}(\omega, {\bf k})}{{\rm i}\omega_n - \omega},
\end{equation}
where $\mathcal{\hat A} = A^{(0)} \hat \tau_0 + A^{(3)} \hat \tau_3 + A^{(1)} \hat \tau_1$ is defined by $\mathcal{\hat A}(\omega, {\bf k}) = -2\mathcal{\hat G}''(\omega + {\rm i}0, {\bf k})$ and also of the spin susceptibility
\begin{equation}
\chi({\rm i}\nu_m,{\bf q}) =-\frac1\pi\int_{-\infty}^\infty
\frac{\chi''(\nu,{\bf q})}{{\rm i}\nu_m-\nu}
\,\mathrm{d}\nu.
\end{equation}
By inserting the spectral representations in Eq.~\eqref{eq:MatsselfE}, the following expression for the imaginary part of the retarded selfenergy on the real axis can be obtained
\begin{equation}\label{eq:realselfE}
\hat \Sigma''(\omega,{\bf k}) = \frac{g^2}{N}\sum_{{\bf q}} \int_{-\infty}^\infty \frac{\mathrm{d}\nu}{2\pi}\,\chi''(\nu,{\bf q}) \, \mathcal{\hat A}(\omega - \nu, {\bf k}-{\bf q})\, [\fermi(\omega - \nu) - \bose(\nu) - 1] \,,
\end{equation}
where $\fermi(\omega)$ and $\bose(\nu)$ are the Fermi and the Bose functions, respectively. The corresponding real part of the selfenergy is calculated via Kramers-Kronig transformation. The evaluation of the expression on the right hand side of Eq.~\eqref{eq:realselfE} represents the most demanding part of the calculation due to the ${\bf q},\omega$-summation for every $E,{\bf k}$-combination. The computational effort can be greatly reduced by using the fact that the expression can be written as a difference of two convolutions of the form
\begin{equation}
X\star Y \bigr |_{\omega,{\bf k}}= \frac{1}{N} \sum_{{\bf q}} \int_{-\infty}^\infty
X(\nu, {\bf q})\, Y(\omega - \nu, {\bf k}-{\bf q})
\,\mathrm{d}\nu\,,
\end{equation}
which can be efficiently evaluated using the fast Fourier transform algorithm. With the above definition, a compact expression for $\Sigma''$ reads
\begin{equation}
\Sigma''^{(\alpha)} = g^2 \left[\chi''\star (\fermi-\tfrac12) A^{(\alpha)} - (\bose+\tfrac12)\,\chi'' \star A^{(\alpha)} 
\right],\qquad \alpha=0,3,1 \,.
\end{equation}

We have solved the selfconsistent equations for the selfenergy iteratively starting with a BCS spectral function. The bare quasiparticles were described by the tight-binding dispersion 
\begin{equation} \label{eq:dispersion}
\epsilon({\bf k})= -2t \lbrack \cos (k_xa) +\cos (k_ya) \rbrack - 4t'\cos (k_xa) \cos (k_ya),
\end{equation}
with $t=380\:\mathrm{meV}$ and $t'=-120\:\mathrm{meV}$. The same form of the model spin susceptibility containing the resonance mode and a continuum as in Refs.~\onlinecite{Casek:2005:PRB,Chaloupka:2007:PRB} was employed. In the notation of Ref.~\onlinecite{Chaloupka:2007:PRB}, we have set $\hbar\omega_0=50\:\mathrm{meV}$,
$\Gamma=20\:\mathrm{meV}$, 
$\xi=2.5a$, 
$\hbar\omega_{\rm C}=400\:\mathrm{meV}$,
$\Gamma_{\rm C}=1000\:\mathrm{meV}$,
$\xi_{\rm C}=1.5a$, 
$b_{\rm M}=2$, 
and $b_{\rm C}=4$.
The main differences with respect to the values of Ref.~\onlinecite{Chaloupka:2007:PRB} are: the energy of the resonance $\hbar\omega_{0}$ is slightly higher ($40\:\mathrm{meV}$ in Ref.~\onlinecite{Chaloupka:2007:PRB}), the ``coherence length" $\xi_{\rm C}$ of the continuum is ca three times higher, and the relative spectral weight of the resonance $b_{\rm M}$ is twice as high. With these values of the parameters, the value of the coupling constant of $g=3\:\mathrm{eV}$ leads to $T_{\rm c}= 133\:\mathrm{K}$ and $\Delta_\mathrm{max}= 45\:\mathrm{meV}$.

After every iteration, the chemical potential $\mu$ was adjusted to keep the electron occupancy at $n_\mathrm{el}=0.85$. Working with the spectral functions, $n_\mathrm{el}$ is evaluated by using the formula 
\begin{equation}
n_\mathrm{el} = 1-\frac{1}{N} \sum_{{\bf k}} \int_{-\infty}^\infty \frac{\mathrm{d}E}{2 \pi} \, A^{(3)}(E,{\bf k})\tanh\frac{\beta E}2 \;.
\end{equation}
For the sampling of the selfenergy, we have used a grid of $128\times128$ points in the Brillouin zone and the energy axis was discretized using $32768$ points covering uniformly the energy range $(-4\:\mathrm{eV},+4\:\mathrm{eV})$.

Finally, after obtaining the selfconsistent solution of the selfenergy equation, the optical conductivity was calculated. The relevant expression reads  
\begin{equation} \label{eq:cond_start}
\sigma_{xx}(\omega) = \frac{{\rm i}e^2 N_p}{\hbar d}
\frac{\Pi_{xx}(\omega) -  K_{xx}}{\omega + {\rm i}0} \;,
\end{equation}
where $d$ is the interplane spacing ($d=9.52$~{\AA} for Hg-1201, $d=11.65$~{\AA} for Y-123), $N_p$ is the~number of CuO$_2$ planes within a unit cell ($N_p = 1$ for Hg-120, $N_p = 2$ for Y-123), $\omega$ is in units of energy and $K_{xx}$ and $\Pi_{xx}(\omega)$ represent the so-called diamagnetic and paramagnetic contributions to the optical conductivity, respectively. The diamagnetic term is given by the following formula (exact)
\begin{equation}
K_{xx} = -\frac{1}{N} \sum_{{\bf k}} 
\frac{\partial^2 \varepsilon_{{\bf k}}}{\partial (k_x a)^2}
\left[
1-\int_{-\infty}^\infty \frac{\mathrm{d}E}{2 \pi} \, A^{(3)}(E,{\bf k})\tanh\frac{\beta E}2
\right].
\end{equation}
The paramagnetic term is given by the retarded current-current correlation function  and can be calculated only approximately. The most frequently used approximation,
employed also in our calculations, completely neglects vertex corrections and leads to the expression for the imaginary part of the response function $\Pi_{xx}(\omega)$ of the form
\begin{equation} \label{eq:cond_fin}
\Pi''_{xx}(\omega) = - \frac{1}{2 N}\sum_{{\bf k}} 
\left(\frac{\partial \varepsilon_{{\bf k}}}{\partial k_x a}\right)^2
\int_{-\infty}^\infty \frac{\mathrm{d}\nu}{2\pi} \,{\rm Tr}\left\lbrace \mathcal{\hat A}(\nu,{\bf k}) \mathcal{\hat A}(\nu + \omega,{\bf k}) \right\rbrace \left\lbrack \fermi(\nu)-\fermi(\nu + \omega) \right\rbrack \,.
\end{equation}
The real part of $\Pi_{xx}(\omega)$ was again obtained by the Kramers-Kronig transformation.

In order to assess the validity of the approximation, we have also evaluated $\Pi_{xx}$ using the gauge invariant approach of Ref.~\onlinecite{Chaloupka:2007:PRB}, where an important class of vertex corrections is included. The corrections modify the spectra of $\Pi''_{xx}$ only slightly, the only significant problem of the approximation being an underestimation of the total spectral weight and $\vert \Pi'_{xx}(0) \vert$ by ca $8\%$, leading to the presence of an unphysical singular component in the normal state. In order to avoid the problem, we have replaced $K_{xx}$ with $\Pi_{xx}'(0)$ when calculating the memory function. For temperatures below $T_{\mathrm{c}}$ we have set $K_{xx}=\Pi_{xx}'(0)[140\mathrm{\,K}]$. The plasma frequency $\omega_{\rm p}$ has been set to $1.91\mathrm{\,eV}$.  

\section{Extended Allen's theory with a gap in the density of states}
\label{sec:2}

The aim of this section is to explore to what extent we can reproduce the experimental data of the in-plane conductivity in the temperature range $T_{\rm c} < T < T^{*}$ using the extended Allen's theory. A special attention is paid to the experimental trend occurring below $T^{\rm ons}$. Our starting point is the fitting procedure developed by Hwang, Sharapov and Carbotte\cite{Sharapov:2005:PRB,Hwang:2011:PRB}. This formalism has been modified so that it is somewhat more rigorous, involves less fitting parameters and also allows one to evaluate all important optical functions.

To describe the model we begin with the boson spectral function $\alpha^2 F$. The Ansatz consists of a single peak described by two parameters $A_{\rm s}$ and $\omega_{\rm s}$
\begin{equation} \label{eq1-eliashberg}
 \alpha^2F(\omega) = \frac{A_{\rm s}\, \omega}{\omega^4 + \omega_{\rm s}^4}, \quad 0 < \omega < \omega_{\rm c},
\end{equation}
an example is shown in Figure~\ref{fig-2-3}~(c). The density of states $N(\omega)$ occurring in the formula for the selfenergy reads
\begin{equation}\label{eq2-dos}
  N(\omega) = 
  \begin{cases}
    1 - h \left ( 1 - \left ( \frac{\omega}{\Delta_{\rm pg}} \right )^2 \right ), &\quad \vert \omega \vert < \Delta_{\rm pg}, \\
    1 + \frac{2}{3} h, &\quad \Delta_{\rm pg} < \vert \omega \vert < 2 \Delta_{\rm pg}, \\
    1, &\quad 2 \Delta_{\rm pg} < \vert \omega \vert,
  \end{cases}
\end{equation}
for an example, see Fig.~\ref{fig-2-3}~(d). Fixing the width of the pseudogap at $\Delta_{\rm pg} = 35$~meV, the end of the ``recovery region''\cite{Hwang:2008:PRL} at $2 \Delta_{\rm pg}$ and conserving the number of states, we have only $2 + 1$ fitting parameters in total --- $A_{\rm s}$, $\omega_{\rm s}$ and the pseudogap depth $h$. The selfenergy $\Sigma$ is calculated within the non--selfconsistent Fock approximation with the gaped density of states $N$. For the imaginary part of the retarded $\Sigma_2$ we have\cite{Sharapov:2005:PRB}
\begin{multline}\label{eq3-sig2}
 \Sigma_2(\omega) = - \pi \int_0^{\infty} {\rm d}\nu \, \alpha^2F(\nu) \Big \lbrace N(\omega - \nu)\left \lbrack \bose(\nu) + 1 + \fermi(\omega - \nu) \right \rbrack  \\ + N(\omega + \nu) \left \lbrack \bose(\nu) + \fermi(\omega + \nu) \right \rbrack \Big \rbrace.
\end{multline}
The real part $\Sigma_1$ is obtained using the Kramers-Kronig relation
\begin{equation}\label{eq4-sig1}
 \Sigma_1(\omega) = -\frac{1}{\pi} \mathcal{P} \int_{-\infty}^{\infty} {\rm d}\nu \, \frac{\Sigma_2(\nu)}{\omega - \nu}.
\end{equation}
The conductivity is calculated using the Allen's theory\cite{Allen:1971:PRB,Allen:2004:condmat}
\begin{align}
 \label{chi}
 \chi(\omega) & = \int_{-\infty}^{\infty} {\rm d}\epsilon \, \frac{{\rm f}(\epsilon) - {\rm f}(\omega + \epsilon)}{\omega - \Sigma(\omega + \epsilon) + \Sigma^{*}(\omega)}, \\ \label{eq6-cond1}
  \sigma(\omega) & = \frac{\imi \epsilon_0}{\hbar}\frac{\omega_{\rm p}^2}{\omega} \chi(\omega).
\end{align}
For the sake of simplicity the integration in \eqref{chi} does not involve the non--constant density of states $N$. The memory function $M$ is obtained as
\begin{equation}
 \label{memory-function}
 M(\omega) = - \omega \left \lbrace \frac{1}{\chi(\omega)} + 1 \right \rbrace.
\end{equation}
Here is the main formal difference between our approach and the one by Hwang~{\it et al}.\cite{Hwang:2011:PRB}, where a simplified expression, obtained by a series expansion of the right hand side of \eqref{chi} in powers of the selfenergy $\Sigma$, is used to calculate $M_2$.

\begin{figure*}[tbp]
\centerline{\includegraphics[width=18cm]{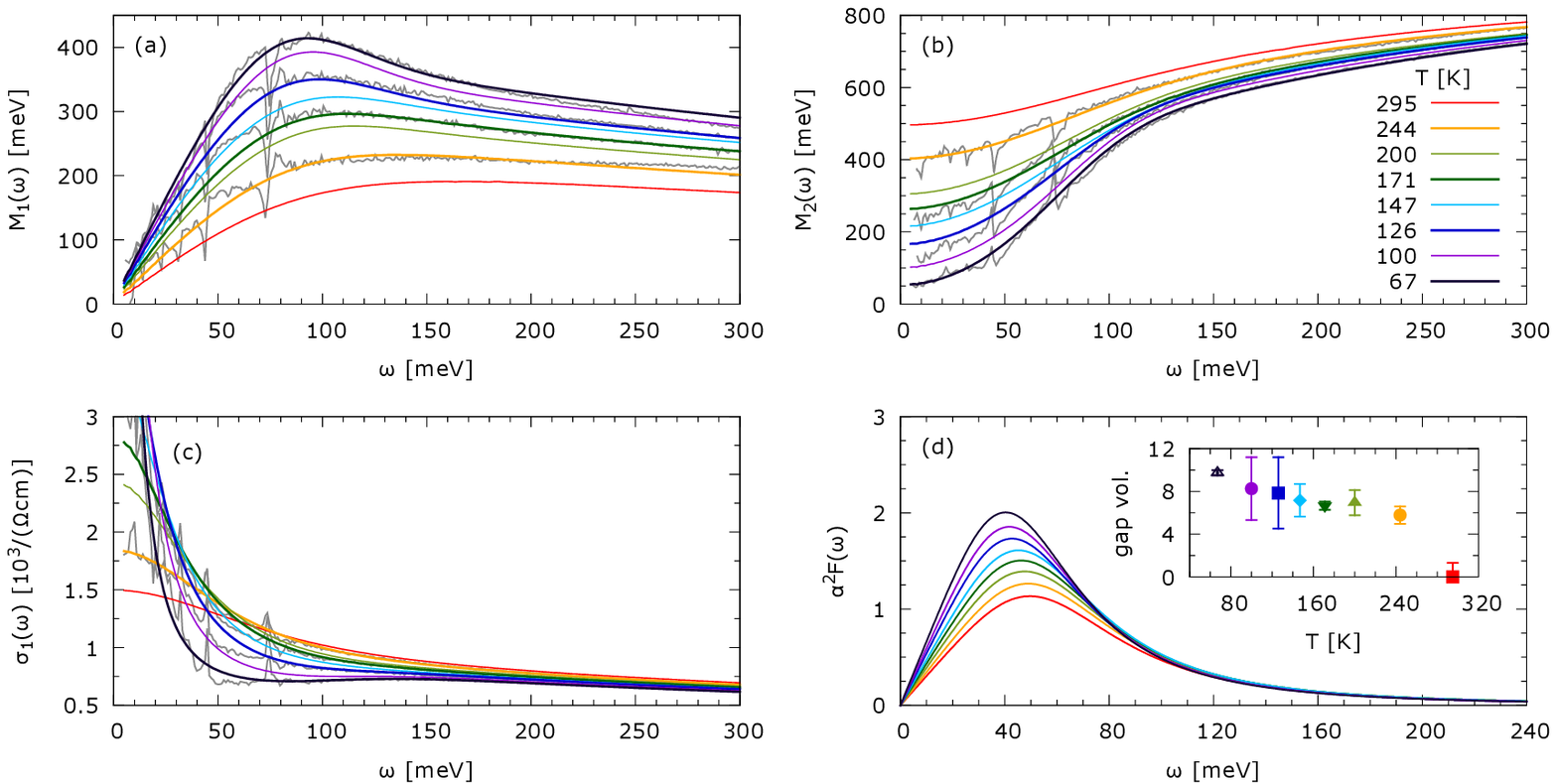}}
\caption{ \label{fig-2-1} (a), (b) Results of the fitting procedure described in the text applied to the experimental data of underdoped Y-123 reported in Ref.~\onlinecite{Hwang:2006:PRB}. The data (gray lines) are presented for several temperatures only. We can see that the quality of the fit is good for all temperatures studied. (c) The resulting spectra of $\sigma_1$ together with selected data. (d) The obtained temperature dependence of the function $\alpha^2F$. We see that with decreasing temperature the maximum of $\alpha^2F$ grows and shifts to lower frequencies. The inset shows the temperature dependence of the gap volume $\frac{2}{3}h\Delta_{\rm pg}$. At 295 K the best fit is with no gap in the density of states and thus the gap volume is zero. Below this temperature the gap volume shows a weak temperature dependence, but with a significant variance of its value.
}
\end{figure*}
\begin{figure*}[tbp]
\centerline{\includegraphics[width=18cm]{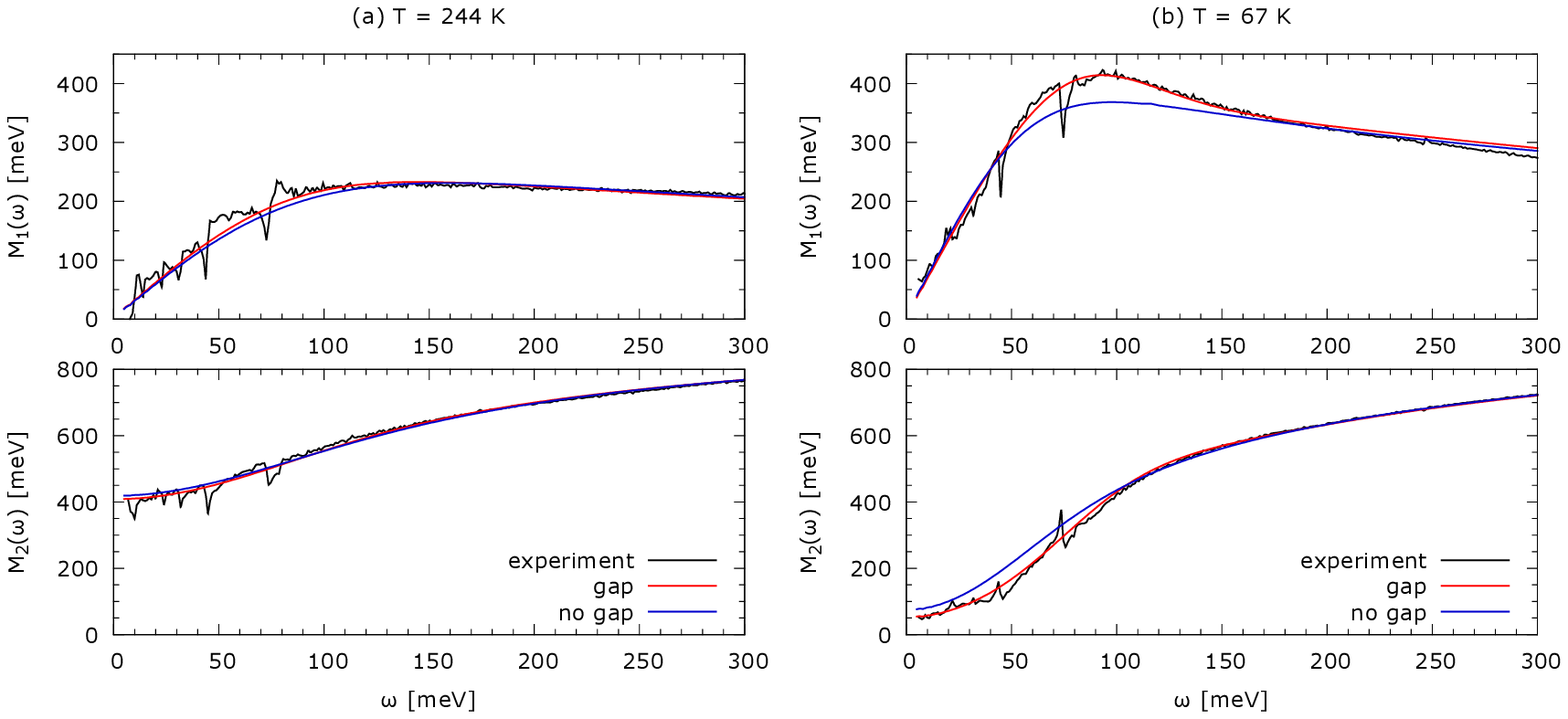}}
\caption{ \label{fig-2-2} Illustration of the impact of the gap in the density of states on the quality of the fit. (a)~Data of UD Y-123 reported in Ref.~\onlinecite{Hwang:2006:PRB} for 244~K together with the results of the fit involving the gap (red line) and with those not involving the gap (blue line). It can be seen that the presence of the gap does not lead to a~significant improvement. (b) The same for 67~K. It can be seen that the role of the gap is essential.
}
\end{figure*}

\begin{figure*}[tbp]
\centerline{\includegraphics[width=18cm]{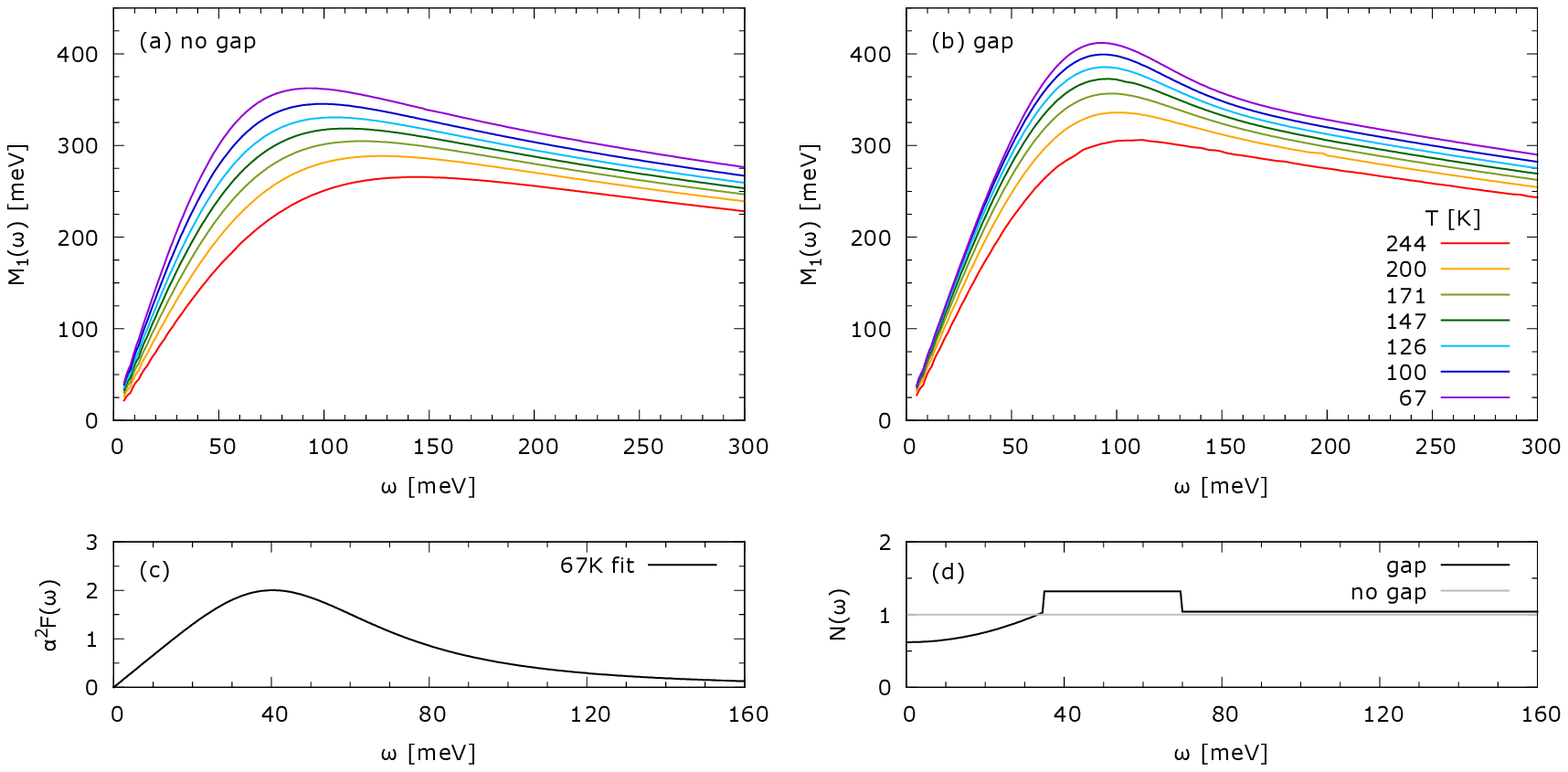}}
\caption{ \label{fig-2-3} Illustration of the role of the gap in the density of states. (a) The temperature dependence of the spectra of $M_1$ calculated using the $\alpha^2F$ from the fit of the 67~K data and no gap in $N$. (b)~The same with the gap included. We clearly see that the characteristic maximum of $M_1$ can be, within the studied model, reproduced by the calculation with the gap in $N$ only. (c)~The function $\alpha^2F$ obtained from the 67~K fit. (d)~Profile of the density of states used in the calculations with (black line) and without (gray line) the gap.
}
\end{figure*}

\begin{figure*}[tbp]
\centerline{\includegraphics[width=18cm]{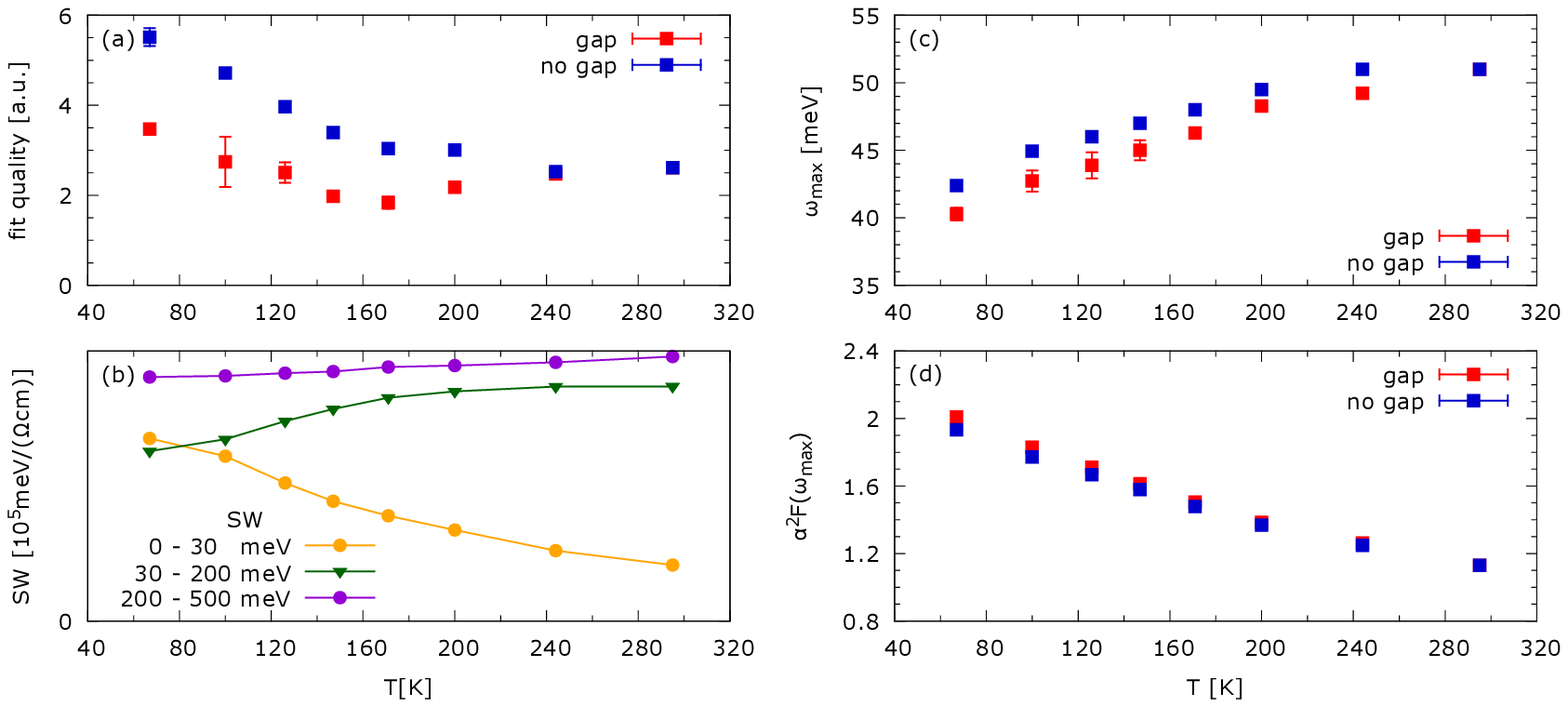}}
\caption{ \label{fig-2-4} (a) Temperature dependence of the quality of the fit of the UD Y-123 data reported in Ref.~\onlinecite{Hwang:2006:PRB} based on the model with the gap in the density of states (red symbols) and that of the model without the gap (blue symbols). It can be seen that for higher temperatures the performance of the two models is comparable. At temperatures closer to $T_{\rm c}$ the model including the gap in $N$ provides significantly better results. (b)~Experimental values (data of Ref.~\onlinecite{Hwang:2006:PRB}) of the optical spectral weight for the three energy ranges indicated 0--30, 30--200 and 200--500 meV. The temperature scale of the beginning of the decrease of the spectral weight in the frequency range 30--200 meV --- already reported in Ref.~\onlinecite{Dubroka:2011:PRL} --- approximately coincides with the onset of the essential importance of the gap in $N$. (c)~The frequency $\omega_{\rm max}$ of the maximum of the boson spectral function as a function of temperature. (d)
~The maximum value $\alpha^2F(\omega_{\rm max})$ of the boson spectral function as a function of temperature. Temperature dependencies of $\omega_{\rm max}$ and $\alpha^2F(\omega_{\rm max})$ are rather monotonic within the range of temperatures studied. We do not observe any change of behavior around 200~K.
}
\end{figure*}

The set of equations \eqref{eq1-eliashberg}--\eqref{memory-function} is iterated for parameters $A_{\rm s}$, $\omega_{\rm s}$ and $h$ using an advanced minimizing procedure --- simulated annealing followed by a simplex method --- to get the smallest value of the norm $\int_{30 \,{\rm meV}}^{400 \,{\rm meV}}\vert M_{\rm num}(\nu) - M_{\rm exp}(\nu) \vert\,{\rm d}\nu$. Here $M_{\rm num}$ is a result of the theoretical calculations described above and $M_{\rm exp}$ is derived from experimental data.

We have applied the described fitting procedure to the experimental data of $\varepsilon_1$ and $\sigma_1$ obtained by Hwang~{\it et al}.\cite{Hwang:2006:PRB} To obtain $M_{\rm exp}$ from the data we have used $\epsilon_\infty = 3.6$. We have also set the plasma frequency $\omega_{\rm p}$ to $2350$ meV (temperature independent), so that $m^*(\omega)/m = M_1(\omega) / \omega + 1 \approx 1$ at $\omega = 1$ eV. Results of the fitting procedure are shown in Figure~\ref{fig-2-1}. The experimental data (gray lines) are presented for several temperatures only. We can see that the quality of the fits is good in the whole range of temperatures. For each temperature we run the procedure several times. The final $\alpha^2F$ is usually almost identical. The maximum of $\alpha^2F$ grows monotonically and shifts to lower frequencies with decreasing temperature, see Fig.~\ref{fig-2-1}~(d). The gap volume $\frac{2}{3}h\Delta_{\rm pg}$, shown in the inset of (d), is zero at 295 K and at lower temperatures 
displays a weak temperature dependence but with a~significant variance of its value.

Closer look at the results for lower and higher temperatures reveals a clear difference in the role of the gap. In Figure \ref{fig-2-2} we compare results obtained by the above described $2 + 1$ parameter model with those obtained using the $2$ parameter model with no gap in the density of states. In (a) we see that for higher temperatures the gap does not play any essential role. It only helps to improve the fit quality slightly. On the other hand in (b) we see that for low temperatures the features in $M_1$ and $M_2$, which are discussed in the main text, can be qualitatively well described only by the model including the gap in the density of states. This is shown in more detail in Fig.~\ref{fig-2-3}, where we show the spectra of $M_1$ calculated using the $\alpha^2F$ from the fit of the 67~K data (a) without and (b) with the gap in the density of states included. The results confirm that within the studied model the discussed low-temperature feature of $M_1$ forms due to the gap in $N$.

In Fig.~\ref{fig-2-4} we review some important characteristics of both types of fits --- with and without the gap in $N$. The measure of the quality of the fit in (a) is given by the integral norm $\vert M_{\rm num}~-~M_{\rm exp} \vert$ described above. We see that for higher temperatures both fits are of a comparable quality. For temperatures closer to $T_{\rm c}$, starting from ca 200~K, better results are achieved by the approach with nonzero gap in $N$. In (b) we show the temperature dependence of the integrated spectral weight, $SW(\omega_1, \omega_2) = \int_{\omega_1}^{\omega_2} {\rm d}\nu \,\sigma_1(\nu)$ of the experimental data from Ref.~\onlinecite{Hwang:2006:PRB}. Comparison of (a) and (b) reveals that the temperature, below which the gap in $N$ is necessary for achieving a good quality of the fit, approximately coincides with the temperature of the onset of the spectral weight shift from the frequency range $30$--$200$~meV to lower frequencies, see the green and the yellow line in Fig.~\ref{fig-2-4}~(b). This temperature scale was already recognized in Ref.~\onlinecite{Dubroka:2011:PRL} and found to be close to $T^{\rm ons}$ of the c-axis data. The temperature dependence of the frequency $\omega_{\rm max}$ of the maximum of the boson spectral density in (c) and that of $\alpha^2F(\omega_{\rm max})$ in (d) are rather monotonic for both methods. There is no apparent change of the basic trend within the range of temperatures studied.

In summary, we have fitted the experimental data reported by Hwang~{\it et al}.\cite{Hwang:2006:PRB} by the model with and without the gap in the density of states. Comparison of the results shows that at high temperatures the model which uses the boson spectral function $\alpha^2F$ and an energy independent density of states is sufficient for a reasonable description of the data. On the other hand, for temperatures closer to $T_{\rm c}$, the gap in $N$ becomes crucial for reproducing the specific features of $M_1$ and $M_2$. The fitting procedure reveals the presence of a characteristic temperature located around $200$~K, below which the presence of the gap in the density of states is necessary. This temperature scale is obviously connected to that described in Ref.~\onlinecite{Dubroka:2011:PRL} (see the corresponding supporting material).

\section{Remarks on purely normal-state interpretations of the gap in the in-plane conductivity}

\begin{figure*}[tbp]
\centerline{\includegraphics[width=18cm]{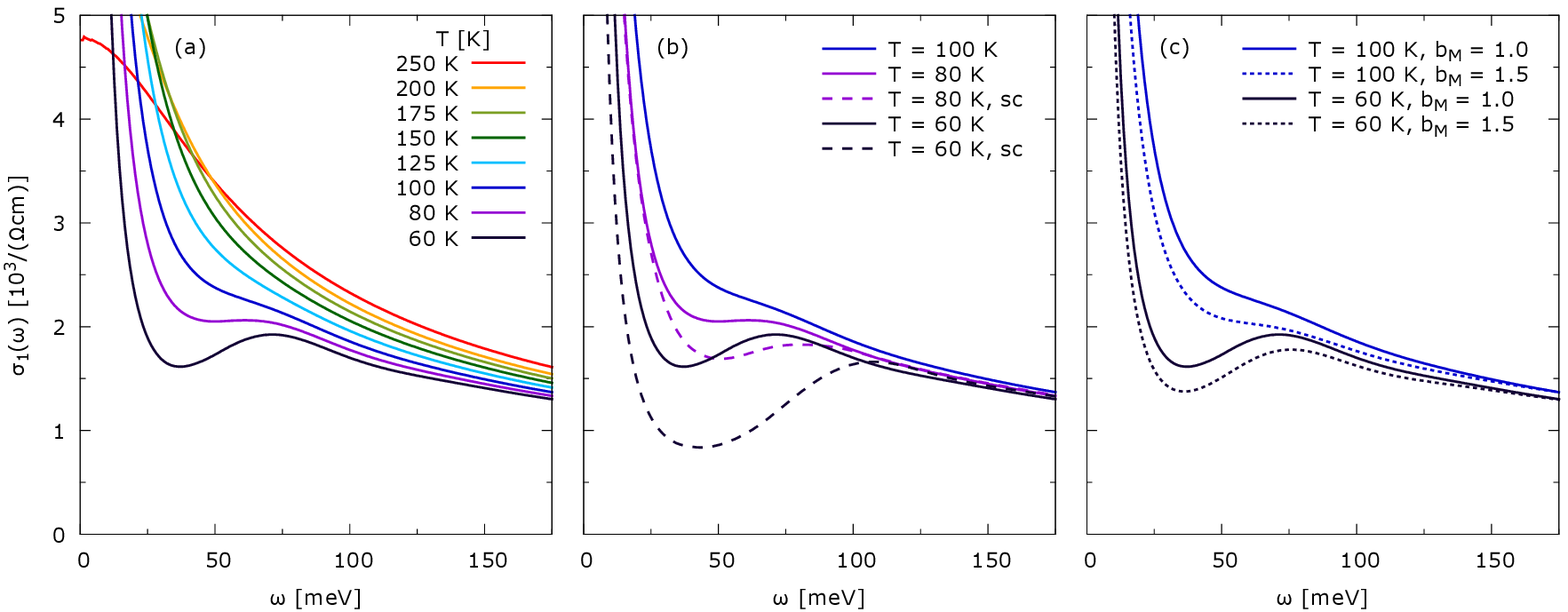}}
\caption{\label{FigureIII1} (a) Normal state temperature dependence of $\sigma_{1}$ calculated as described in the text. (b) Selected normal state spectra from (a) ($100\mathrm{\,K}$, $80\mathrm{\,K}$, $60\mathrm{\,K}$) and the $80\mathrm{\,K}$ and $60\mathrm{\,K}$ superconducting state spectra. The difference between the normal state gap feature due to the temperature dependence of the Bose and Fermi factors and the superconducting state gap feature due to the opening of the superconducting gap can be clearly seen. (c) Two normal state spectra from (a) ($100\mathrm{\,K}$ and $60\mathrm{\,K}$) and the $100\mathrm{\,K}$ and $60\mathrm{\,K}$ normal state spectra calculated with a higher value of the spectral weight of the resonance of $b_{\rm M}=1.5$. A deepening of the gap due to the increase of $b_{\rm M}$ can be clearly seen.}
\end{figure*}

First, a gap in the in-plane conductivity develops with decreasing temperature naturally, even in the absence of a DOS gap, due to the temperature dependence of the Bose and Fermi factors. Its scale is determined by the characteristic boson energy and it is typically narrower than in the data. An example of a standard normal state temperature dependence of $\sigma_{1}$ is shown in Fig.~\ref{FigureIII1}~(a). The spectra have been calculated using the Eliashberg formalism described in Sec.~\ref{sec:1} and the same values of the input parameters as in Ref.~\onlinecite{Chaloupka:2007:PRB} except for $t=0.38\mathrm{\,eV}$, $t'=-0.120\mathrm{\,eV}$, $n=0.85$ and $g=3.0\mathrm{\,eV}$, for definitions, see Refs.~\onlinecite{Chaloupka:2007:PRB} and \onlinecite{Casek:2005:PRB}. The onset of superconductivity leads to a different shape of the gap. This is illustrated in Fig.~\ref{FigureIII1}~(b) showing the $100\mathrm{\,K}$, $80\mathrm{\,K}$ and $60\mathrm{\,K}$ normal state spectra from (a), and the $80\mathrm{\,K}$ 
and $60\mathrm{\,K}$ superconducting state spectra. For the present values of the input parameters, $T_{\mathrm{c}}=90\mathrm{\,K}$.

\begin{figure*}[tbp]
\centerline{\includegraphics[width=18cm]{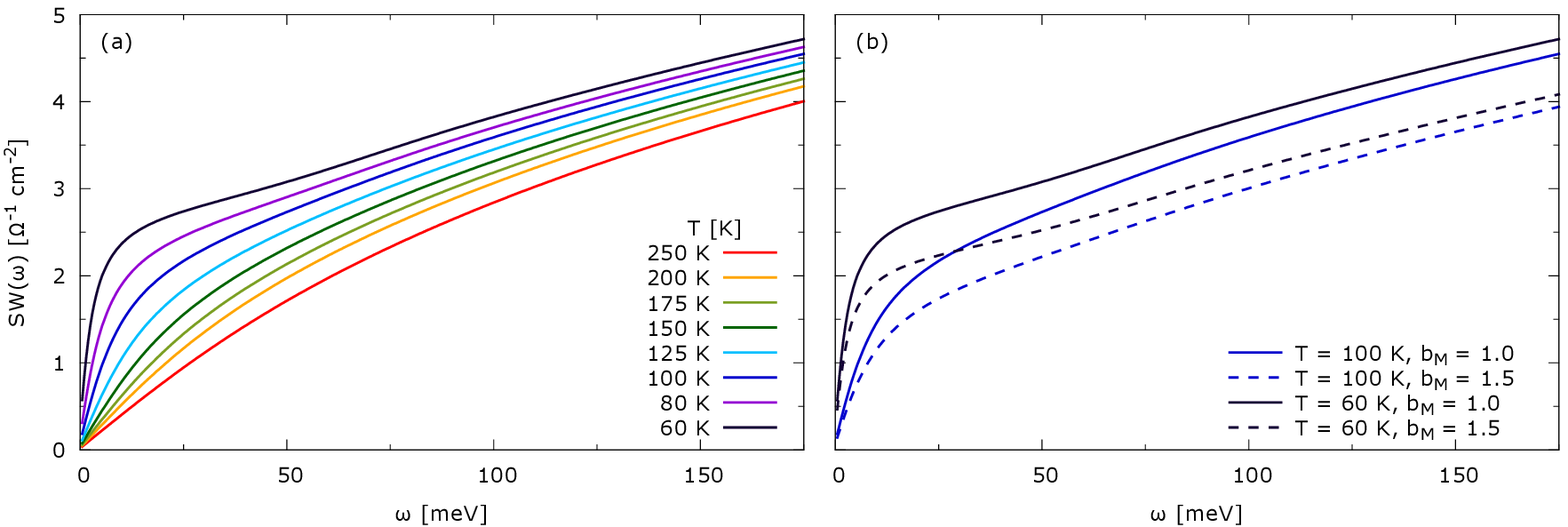}}
\caption{\label{FigureIII2} (a) Normal state temperature dependence of the frequency dependent spectral weight $SW(\omega)$ defined in the text corresponding to the temperature dependence of the conductivity  shown in Fig.~\ref{FigureIII1}~(a). (b) The spectral weight $SW(\omega)$ for two values of temperature and for two values of the spectral weight $b_{\rm M}$ of the low energy component of the bosonic spectral function. It can be seen that $SW(\omega)$ decreases with increasing $b_{\rm M}$.}
\end{figure*}

Second, a deepening of the gap for a fixed temperature can be achieved by increasing the spectral weight of the low-energy component of the bosonic spectral function ($\alpha^{2}F$ in the Allen's theory, $\chi''({\mathbf q},\omega)$ in realistic models involving spin fluctuations). The latter increase, however, also leads to a reduction of the total optical spectral weight in the infrared, not occurring in the data. This problem is illustrated in Figures~\ref{FigureIII1}~(c) and \ref{FigureIII2}. Figure~\ref{FigureIII1}~(c) shows the $100\mathrm{\,K}$ and $60\mathrm{\,K}$ normal state spectra from (a) and the $100\mathrm{\,K}$ and $60\mathrm{\,K}$ normal state spectra calculated with a higher value of the spectral weight of the resonance of $b_{\rm M}=1.5$ ($b_{\rm C}$ is reduced accordingly, $b_{\rm C}=3.5$). The deepening of the gap can be clearly seen. Figure~\ref{FigureIII2}~(a) shows the calculated normal state temperature dependence of the spectral weight $SW(\omega)=\int_{0}^{\omega}\sigma_{1}(\omega')
d\omega'$  corresponding to the spectra of Fig.~\ref{FigureIII1}~(a), that is qualitatively consistent with the experimental one shown in Fig.~4 of Ref.~\onlinecite{Hwang:2006:PRB}. With decreasing temperature, $SW(\omega)$ at low frequencies {\it increases}. Figure~\ref{FigureIII2}~(b) shows that the increase of $b_{\rm M}$, for a fixed temperature, leads to the opposite trend: with increasing $b_{\rm M}$ the spectral weight at low frequencies {\it decreases}. It is thus unlikely that the formation of the gap in $\sigma_{1}$ could be described as being solely due to a specific temperature dependence of the low energy component of $\alpha^{2}F$ or $\chi''({\mathbf q},\omega)$.  

\section{Simplified calculations involving the superconducting gap --- ``Hybrid model''}

Here we provide a detailed information about our study of the impact of the gap of superconducting nature on the in-plane conductivity in the temperature range $T_{\rm c} < T < T^{*}$. Our approximate retarded Green's function is given by
\begin{equation}
  \hat G(\omega, {\bf k}) = \frac{(\omega - \Sigma(\omega))\hat \tau_0 + (\epsilon({\bf k}) - \mu)\hat\tau_3 + \phi(\omega,{\bf k})\hat\tau_1}{(\omega - \Sigma(\omega))^2 - (\epsilon({\bf k}) - \mu)^2 - \phi^2(\omega,{\bf k})}.
\end{equation}
Here $\Sigma(\omega)$ is the quasiparticle selfenergy, $\epsilon({\bf k})$ the dispersion relation, $\mu$ the chemical potential, and $\phi(\omega,{\bf k})$ reads
\begin{equation}
  \phi(\omega,{\bf k}) = \left (1 - \frac{\Sigma(\omega)}{\omega} \right) \Delta_{\rm sc}({\bf k}),
\end{equation}
where $\Delta_{\rm sc}(\omega)$ is the superconducting gap of d-wave symmetry
\begin{equation}
  \Delta_{\rm sc}({\bf k}) = \frac{\Delta_0}{2} \left \lbrack \cos(k_x a) - \cos(k_y a) \right \rbrack.
\end{equation}
This definition of $\phi(\omega,{\bf k})$ ensures that the gap in the upper diagonal component of the quasiparticle spectral function $\mathcal{\hat A}(\omega,{\bf k})$ defined by Eq.~\eqref{eq:GreenSpectral} is exactly $\Delta_{\rm sc}({\bf k})$ for $\epsilon({\bf k}) = \mu$. In order to calculate the in-plane conductivity we decompose it into the regular part $\sigma_{xx}^{(r)}$ and the singular part $\sigma_{xx}^{(s)}$: $\sigma_{xx}(\omega) = \sigma_{xx}^{(r)}(\omega) + \sigma_{xx}^{(s)}(\omega)$. The regular part is defined by
\begin{equation}
 \sigma_{xx}^{(r)}(\omega) = \frac{{\rm i}e^2 N_p}{d \hbar \omega} \left \lbrack \Pi_{xx}(\omega) - {\rm Re}\lbrace \Pi_{xx}(0) \rbrace \right \rbrack\,,
\end{equation}
where $\Pi_{xx}(\omega)$ is given by Eq.~\eqref{eq:cond_fin}. The singular can be expressed as
\begin{equation}
 \sigma_{xx}^{(s)}(\omega) = \frac{{\rm i}\epsilon_0 \, \omega^2_{\rm pl,sc}}{\omega + {\rm i}0}\,,
\end{equation}
where $\omega_{\rm pl,sc}$ is the plasma frequency of the superfluid,
\begin{equation}
 \omega^2_{\rm pl,sc} = \omega^2_{\rm pl} - \frac{2}{\pi \epsilon_0}\int_{0_+}^{\infty} {\rm d}\omega \,\sigma_{xx1}(\omega)
\end{equation}
and $\omega_{\rm pl}$ is calculated from the normal state spectra by 
\begin{equation}
  \omega_{\rm pl}^2 = \frac{2}{\pi\epsilon_0}\int_0^{\infty} {\rm d}\omega \, \sigma_{xx 1}(\omega)\,.
\end{equation}

We employ this formalism in two different situations. First, we study the impact of the superconducting gap $\Delta_{\rm sc}$ on features of $\sigma_1$ due to the superconductivity unrelated gap in the antinodal region. In order to do this we use $\Sigma(\omega)$ given by \eqref{eq3-sig2} and \eqref{eq4-sig1}, with $\alpha^2F(\omega)$ from our fits of the 100~K and 67~K data reported in Ref.~\onlinecite{Hwang:2006:PRB} and with the value of the gap depth of $h = 1$. We use the dispersion relation given by Eq. \eqref{eq:dispersion}, with $t = 250$~meV and $t' = -100$~meV. The chemical potential $\mu$ is set to $-350$~meV. Our numerical calculations show that the incorporation of the superconducting gap $\Delta_{\rm sc}$ causes a shift of the studied features of $\sigma_1$, $M_1$ and $M_2$, as discussed in the main text, see Fig.~1 (g), (h) and (i) there.

Second, we calculate the spectra of $\sigma_1$ using the properties of $\mathcal{A}(\omega,{\bf k})$ reported by Reber~{\it et al}., see Refs.~\onlinecite{Reber:2012:NaturePhys}~and~\onlinecite{Reber:2013:PRB}. We use a form of selfenergy which is similar to the one used there (see page 5 of their supplementary information). The imaginary part $\Sigma_2(\omega)$ is step-like with a temperature dependent low energy part $\Gamma(T)$:
\begin{equation}
\Sigma_2(\omega,T) = 
 \begin{cases}
  \Gamma(T),& \vert \omega \vert < \Omega_{\Gamma}, \\
  \Gamma_{\infty},& \Omega_{\Gamma} \leq \vert \omega \vert.
 \end{cases}
\end{equation}
Temperature dependence of $\Gamma(T)$ is taken from Fig.~4~(a) of Ref.~\onlinecite{Reber:2012:NaturePhys}, $\Omega_{\Gamma} = 70$~meV and $\Gamma_\infty = -300$~meV. Real part of the selfenergy $\Sigma_1(\omega)$ is calculated using the Kramers-Kronig relation. We set the value of $\Delta_{\rm sc}$ to 40~meV, consistent with Fig.~4~(a) of Ref.~\onlinecite{Reber:2012:NaturePhys}, with no temperature dependence for simplicity. We use the same dispersion relation and the same values of its parameters as in the previous paragraph. Our results are presented in Fig.~3 of the main text. To illustrate the importance of the anomalous part of the Green's function we also present spectra of $\sigma_1(\omega)$ calculated using the diagonal parts of $\mathcal{\hat A}(\omega,{\bf k})$ only.